\documentclass[hyper]{JHEP} 

\usepackage{epsfig}

\newcommand\fverb{\setbox\pippobox=\hbox\bgroup\verb}

\newcommand\fverbdo{\egroup\medskip\noindent%

            \fbox{\unhbox\pippobox}\ }

\newcommand\fverbit{\egroup\item[\fbox{\unhbox\pippobox}]}

\newbox\pippobox

\title{Note About Canonical Formalism for Normalized Gravity
And Vacuum Energy Sequestering Model}
\author{J. Kluso\v{n}\\
Department of
Theoretical Physics and Astrophysics\\
Faculty of Science, Masaryk University\\
Kotl\'{a}\v{r}sk\'{a} 2, 611 37, Brno\\
Czech Republic\\
E-mail: \email{klu@physics.muni.cz}}
\preprint{}

 \abstract{This short note is devoted to the Hamiltonian
 analysis of the normalized general relativity and recently proposed
 model of vacuum energy sequestering. The common property of these models is
 the presence of the global variables.  We discuss the meaning of these
  global variables in the context of the
 canonical formalism and argue that their presence lead to the
 non-local form of the Hamiltonian constraint. }
 \keywords{Hamiltonian Formalism, Normalized General Relativity}

\def\be{\begin{equation}}

\def\ee{\end{equation}}

\def\bea{\begin{eqnarray}}

\def\eea{\end{eqnarray}}

\def\mH{\mathcal{H}}

\def\bx{\mathbf{x}}
\def\by{\mathbf{y}}

\newcommand{\hg}{\hat{g}}

\newcommand{\mG}{\mathcal{G}}

\newcommand{\bT}{\mathbf{T}}

\newcommand{\mL}{\mathcal{L}}

\def\pb #1{\left\{#1\right\}}

\begin{document}
\section{Introduction and Summary}\label{first}
One of the most serious mystery of today's physics is the value of
the cosmological constant \cite{Weinberg:1988cp}. There are many
proposals how to explain the small values of given constant, for
review of some of the most popular ones, see \cite{Copeland:2006wr}.
Another less known  model  that explains the origin of cosmological
constant is unimodular gravity
\cite{Ellis:2010uc,Ellis:2013uxa,Padilla:2014yea,Gao:2014nia,
Jain:2012gc,Smolin:2010iq,Smolin:2009ti,Shaposhnikov:2008xb,
Alvarez:2005iy,Finkelstein:2000pg,Barcelo:2014mua,Barcelo:2014qva}.
In the framework of unimodular gravity
 the cosmological constant $\Lambda$ is interpreted
as an arbitrary constant of integration. However even if the
cosmological constant is now determined by initial conditions we
still do not know how to control its value.

Due to the fact that the value of the cosmological constant is very
small one can wonder why it is not exactly equal to zero. In other
words it would be interesting to find some \emph{symmetry principle}
that enforces the cosmological constant to be zero. Clearly the
problem why the cosmological constant has so tiny value remains.

However the requirement that the cosmological constant is zero has
remarkable consequence for the structure of the theory. In fact,
this requirement can be enforced when we demand that the Lagrangian
density is invariant under shift $\mL\rightarrow
\mL+\mathrm{const}$. Clearly when the theory obeys this symmetry the
cosmological constant is trivial. Then the requirement that the
theory is invariant under constant shift of the Lagrangian leads to
so called \emph{normalized} Einstein-Hilbert action
\begin{equation}
I=\frac{\int d^4x\sqrt{-\hg}(\frac{1}{16\pi
G}{}^{(4)}R+\mL_m)}{\epsilon \int d^4x \sqrt{-\hg}} \ ,
\end{equation}
where $\epsilon$ is the parameter of the mass dimension $M^4$ in
order to make given action dimensionless. The above action was
firstly introduced by Tseytlin \cite{Tseytlin:1990hn} where he
considered this theory as an effective low energy limit of some
duality symmetric closed string theory. On the other hand given
theory can be formulated without this assumption, see for example
\cite{Davidson:2014vda,Davidson:2009mp}.

Finally recently new proposal how to solve cosmological constant
problem was proposed in
\cite{Kaloper:2013zca,Kaloper:2014dqa,Kaloper:2014fca}. It is based
on the mechanism that ensures that all the vacuum energy from a
matter sector is sequestered from gravity. The idea it to make all
scales in this matter sector functionals of the four volume element
of the Universe. It turns out that the mechanism is minimal
modification of general relativity when we add to the action
auxiliary fields with an extra term that is not integrated over.
Explicitly, this mechanism is described by the action
\begin{equation}\label{actionKal}
S=\int d^4x \sqrt{-\hg}\left[\frac{1}{16\pi G}{}^{(4)}R
-\Lambda-\lambda^4 \mL(\lambda^{-2}\hg^{\mu\nu},\Phi)\right]+
\sigma\left(\frac{\Lambda}{\lambda^4\mu^4}\right) \ ,
\end{equation}
when the matter with Lagrangian density $\mL$ couples minimally to
the rescaled metric $\tilde{g}_{\mu\nu}=\lambda^2 \hg_{\mu\nu}$. In
this proposal $\Lambda$ and $\lambda$ are dynamical variables
without any local dynamics.

It is important to stress that they are not conventional auxiliary
fields where we can presume that they vary over the space-time
but with no kinetic terms in the Lagrangian.
  Instead the variables used in the models above
   are global variables with no dependence of
space-time coordinates. However this fact implies that even if given
theory can be interpreted as a minimal modification of the general
relativity the presence of the global variables has crucial impact
on the Hamiltonian formulation. More precisely, the fact that these
variables are global one implies that it is not possible to define
their conjugate momenta. This is crucial difference from the case of
auxiliary fields that appear in the Lagrangian density so that we
can introduce their conjugate momenta as the primary constraints of
the theory. We argue that in case of the global variables we should
proceed in different way. First of all we introduce  Hamiltonian for
all dynamical fields that appear in the Lagrangian.
 Using this Hamiltonian
and corresponding canonical variables we perform the variation of
the action which leads to the equations of motion that can be
expressed using conventional Poisson brackets. On the other hand the
variation of the action with respect to the global variables leads to
the set of relations between canonical variables. Note that by
nature of these global variables these relations include the
integration over the space and time. As the final step we
include these relations  to the equations of motion for the canonical variables. Say
differently, if we express the equations of motion in terms of
Poisson brackets we find that these global relations should be
inserted to these Poisson brackets \emph{after} their explicit
evaluations. We demonstrate that this procedure is generally valid
even for the system with constraints.

Following this general discussion we  determine corresponding
Hamiltonian structure for normalized general relativity and for the
action (\ref{actionKal}). We firstly calculate the Poisson brackets
between canonical variables and then insert the values of the global
variables to them. Then it is clear that  if we found that the
Poisson bracket vanishes on the constraint surface it will vanish on
the constraint surface   even with the explicit values of the global
variables included. In other words the constraint structure of all
these theories is the same as in case of standard general relativity
\cite{Isham:1984sb,Isham:1984rz}. Of course, the problem is that the
equations of motion for the canonical variables are non-local since
they depend on the integrals of the canonical variables defined over
the whole space-time. But when we presume that these global
constraints can be fixed in some way we obtain well defined system
with the clear constraint structure.

This paper is organized as follows. In the next section
(\ref{second}) we perform canonical analysis of general systems with
global variables. Using this formalism we proceed to the Hamiltonian
analysis of the normalized general relativity in section
(\ref{third}). Finally in section (\ref{fourth}) we perform
Hamiltonian analysis of the model proposed in
\cite{Kaloper:2013zca,Kaloper:2014dqa,Kaloper:2014fca}.
\section{Hamiltonian Analysis for Systems with Global Variables}
 \label{second}
Let us consider an action in the following form
\begin{equation}
S=\int dt L(q,\dot{q},\Lambda)+f(\Lambda)
 \ ,
 \end{equation}
where $f$ is some function that depends on global variable
$\Lambda$ and where $q$ generally means
set of dynamical variables.  The equations of motion for $q$ and $\Lambda$ that arise
by variation of the action with respect to $q$ and $\Lambda$ take
the form
\begin{equation}\label{eqq}
\frac{\delta L}{\delta q}-\frac{d}{dt}\left(\frac{\delta L}{\delta
\dot{q}}\right)=0 \ , \quad  \int dt \frac{\delta L}{\delta
\Lambda}+\frac{df}{d\Lambda}=0 \ .
\end{equation}
Let us  presume that the last equation can be solved for $\Lambda$
as a function of $q,\dot{q}$ so that $\Lambda=\Lambda(
q',\dot{q}')$, where $q'\equiv q(t'), \dot{q}'=\dot{g}(t')$ and
where generally $\Lambda$ depends on $q',\dot{q}'$ through the
integral over $t'$. Then inserting this value of $\Lambda$ into the
first equation of (\ref{eqq}) we derive the equation of motion for
$q$ that is manifestly non-local.

Let us now try to formulate Hamiltonian formalism for given system.
We define the canonical momentum $p$ and the Hamiltonian $H$  in the
usual manner as $p=\frac{\delta L}{\delta \dot{q}}, H=p\dot{q}-L$.
Note that due to the fact that $\Lambda$ is global variable it does
not make sense to define corresponding conjugate momentum.
This can be also seen from the fact that this variable is not included
in the Lagrangian of the theory.
 It is
instructive to compare this situation with the possibility when
$\Lambda$ depends on $t$ at least in principle. In this case  we can
introduce the conjugate momentum that, due to the fact that the
Lagrangian does not depend on time derivative of $\Lambda$, is the
primary constraint of the theory. The presence of such primary
constraint would lead to the emergence of the secondary constraint
which will be manifestly local.


Returning to our case we formulate the action using the canonical
variables $p$ and $q$ so that it has the form
\begin{equation}
S=\int dt (p\dot{q}-H(p,q,\Lambda))+f(\Lambda) \ .
\end{equation}
The variation of the action has the form
\begin{eqnarray}
\delta S=\int dt \left(\delta p \dot{q}-\dot{p}\delta q-
\frac{\delta H}{\delta p}\delta p-\frac{\delta H}{\delta q}\delta
q-\frac{\delta
H}{\delta \Lambda}\delta \Lambda\right)+\frac{df}{d\Lambda}\delta\Lambda=0 \nonumber \\
\end{eqnarray}
so that we derive following equations of motion
\begin{eqnarray}
& &\dot{q}=\frac{\delta H(p,\Lambda)}{\delta p} \ , \nonumber \\
& &\dot{p}=-\frac{ \delta H(p,\Lambda)}{\delta q} \ , \nonumber \\
& &\int dt \frac{\delta H}{\delta
\Lambda}=\frac{df}{d\Lambda} \  . \nonumber \\
\end{eqnarray}
Let us presume  that the last equation can be solved for $\Lambda$,
at least in principle, so that $\Lambda=\Lambda(p',q')$ Then we can
insert this value to the first two equations and we derive
\begin{eqnarray}
\dot{q}(t)=\frac{\delta H(q,p,\Lambda(q',p'))} {\delta p} \ , \quad
\dot{p}=-\frac{\delta H(q,p,\Lambda(q',p'))}{\delta q} \ .\nonumber
\\
\end{eqnarray}
Note that this dependence of $\Lambda$
was introduced \emph{after} the variation with respect to $q$ and
$p$ was performed. Finally note the well known relation between
variation of the Hamiltonian with respect to $q,p$ and the Poisson
brackets so that  we can rewrite these equations of motion into the
form
\begin{eqnarray}
\dot{q}=\pb{q,H}(\Lambda(q',p')) \ , \nonumber \\
\dot{p}=\pb{p,H}(\Lambda(q',p')) \ , \nonumber \\
\end{eqnarray}
where again the explicit dependence of $\Lambda$ on $q',p'$ is
introduced \emph{after} the calculation of the Poisson brackets.
Clearly this procedure can be extended to any phase space functions
$f(p,q)$.

On the other hand let us consider the more general case of systems
with constraints. Let us presume that we have set of primary
constraints $\phi_n(q,p,\Lambda)\approx 0$ that follow from the form
of the Lagrangian. Note that generally $\phi_n$ depend on $\Lambda$.
 Then that extended Hamiltonian takes the form
\begin{equation}
H_E=H+\lambda^n\phi_n(q,p,\Lambda)
\end{equation}
Now the variation principle takes the form
\begin{eqnarray}
\delta S&=& \int dt \left(\delta p \dot{q}-\dot{p}\delta q
-\frac{\delta H}{\delta p}\delta p-\frac{\delta H}{\delta q}\delta q
-\frac{\delta
H}{\delta \Lambda}\delta \Lambda-\right.\nonumber \\
&- &\left. \lambda^n\frac{\delta \phi_n}{\delta q}\delta
q-\lambda^n\frac{\delta \phi_n}{\delta p}\delta
p-\lambda^n\frac{\delta \phi_n}{\delta \Lambda}\delta
\Lambda\right)+\frac{d f}{d \Lambda}\delta
\Lambda=0 \nonumber \\
\end{eqnarray}
where we ignored the variation of the action with respect to
$\lambda$ that implies the constraint $\phi_n\approx 0$. From this
variation we derive the equation of motion
\begin{eqnarray}
& &\dot{q}=\frac{\delta H}{\delta p}+\lambda^n\frac{\delta
\phi_n}{\delta p} \ ,  \nonumber \\
& &\dot{p}=-\frac{\delta H}{\delta q}-\lambda^n\frac{\delta \phi_n}
{\delta q} \ , \nonumber \\
& &\int dt \left[\frac{\delta H}{\delta \Lambda}+\lambda^n \frac{\delta
\phi_n}{\delta\Lambda}\right]=\frac{d f}{d \Lambda} \ .
\nonumber \\
\end{eqnarray}
 Let us for simplicity presume that $\phi_n$
does not depend on $\Lambda$. Then from the last equation we can
find $\Lambda$, at least in principle.
 Further the requirement of
the preservation of the constraints $\phi_n$ is equal to
\begin{equation}
\frac{d\phi_n}{dt}=\pb{\phi_n,H}(\Lambda)+ \lambda^m
\pb{\phi_n,\phi_m}=0 \ .
\end{equation}
If $\pb{\phi_n,\phi_m}$ is regular matrix then this equation can be
solved for $\lambda^m$. On the other hand if $\pb{\phi_n,\phi_m}$ is
zero we determine another secondary constraints
$\psi_i(p,q,\Lambda)$. We should include these constraints into the
definition of the Hamiltonian so that we find the most general case
where we have an extended  Hamiltonian with collection of all
constraints included where the constraints generally depend on
$\Lambda$. Explicitly,  let us denote $\Phi_I$ as the collection of
all constraints, primary, secondary and so on so that the total
Hamiltonian has the form \footnote{For review of the Hamiltonian
analysis of constrained systems, see
  \cite{Govaerts:2002fq,Henneaux:1992ig}.}
\begin{equation}
H_T=H+\lambda^I\Phi_I \
\end{equation}
so that the extended action has the form
\begin{equation}\label{actPhi}
S=\int dt (p\dot{q}-H_T)+f(\Lambda)= \int dt
(p\dot{q}-H-\lambda^I\Phi_I)+f(\Lambda) \ .
\end{equation}
Then the requirement of the preservation of
all constraints has the form
\begin{equation}\label{consPhiI}
\frac{d\Phi_I}{dt}=\pb{\Phi_I,H}(\Lambda) +\lambda^J
\pb{\Phi_I,\Phi_J}=0 \ .
\end{equation}
 Let us presume that
these equations can be   solved for
$\lambda^I=\lambda^I(p,q,\Lambda)$. Then inserting these
values of $\lambda^I$ into the "equation of motion"
for $\Lambda$ that follows from  (\ref{actPhi})
\begin{equation}\label{eqlambda}
\int dt\left(\frac{\delta H}{\delta \Lambda}+\lambda^I \frac{\delta
\Phi_I}{\delta \Lambda}\right)=\frac{df}{d\Lambda}
\end{equation}
we find the equation for $\Lambda$ that can be solved for the
canonical variables, at least in principle. Finally plugging this
value of $\Lambda$ into $H$ and $\Phi^I$ we find non-local form of
the Hamiltonian for given system. However it is important to stress
that the constraint structure of given system does not depend on the
explicit dependence of $\Lambda$ on the canonical variables as
follows from the fact that we calculate the Poisson brackets with
fixed $\Lambda$.

Let us also discuss the possibility that some of the Lagrange multipliers
 $\lambda$ were not
fixed by equation  (\ref{consPhiI}) which would imply that corresponding
constraints are the first class. Then we wound find that $\Lambda$
depends on non specified gauge parameters. On the other hand we can
fix this gauge freedom by introducing corresponding set of gauge
fixing functions that we denote as $\mG_\alpha$ so that the extended
action has the form
\begin{equation}
S=\int dt (p\dot{q}-H-\Lambda^I\Phi_I-\omega^\alpha
\mG_\alpha)+f(\Lambda) \ .
\end{equation}
Then by definition of the gauge fixing functions we find that all
Lagrange multipliers $\Lambda^I,\omega_\alpha$ are determined by the
requirement of the preservation of the constraints
$\phi_I,\mG_\alpha$ during the time evolution of the system. Finally
we determine $\Lambda$ from (\ref{eqlambda}) with additional term
$\omega^\alpha\frac{\delta \mG_\alpha}{\delta \Lambda}$.

In the next two sections we apply this general analysis to the cases
of two models introduced in the introduction section.
%

\section{Normalized General Relativity}\label{third}
The normalized Einstein-Hilbert action has the form
\begin{equation}
S=\frac{1}{\epsilon \int d^4x\sqrt{-\hg} } \int d^4x\sqrt{-\hg}
\left(\frac{1}{16\pi G}{}^{(4)}R+\mL_{matter}\right) \ ,
\end{equation}
where $\epsilon$ is constant with dimension $[\epsilon]=M^{4}$. To
proceed to the Hamiltonian formalism we introduce two auxiliary
 global variables $A$ and $B$ and rewrite the action into
the form
\begin{equation}\label{SAB}
S=\frac{1}{A}\int d^4x\sqrt{-\hg} \left(\frac{1}{16\pi
G}{}^{(4)}R+\mL_{matter})+B(A-\epsilon \int d^4x \sqrt{-\hg}\right)
\ .
\end{equation}
The equations of motion that follow from this action have the form
\begin{eqnarray}
& &A-\epsilon \int d^4x \sqrt{-\hg}=0 \ , \nonumber \\
&-&\frac{1}{A^2} \int d^4x\sqrt{-\hg} (\frac{1}{16\pi
G}{}^{(4)}R+\mL_{matter})+B=0 \ , \nonumber \\
& &\frac{1}{A}[\frac{1}{16\pi
G}(R_{\mu\nu}-\frac{1}{2}\hg_{\mu\nu}R)-T_{\mu\nu}]+\frac{1}{2}B\epsilon
\hg_{\mu\nu}=0 \ . \nonumber \\
\end{eqnarray}
From the first equation we find $A=\epsilon \int d^4x \sqrt{-\hg}$
so that the second equation gives
\begin{equation}
B=\frac{1}{(\epsilon \int d^4x \sqrt{-\hg})^2} \int d^4x\sqrt{-\hg}
(\frac{1}{16\pi G}{}^{(4)}R+\mL_{matter}) \ .
\end{equation}
Inserting these two expressions to the equation of motion for
$\hg_{\mu\nu}$ we finally obtain
\begin{eqnarray}
& &\frac{1}{16\pi G}(R_{\mu\nu}-\frac{1}{2}\hg_{\mu\nu}R)
-T_{\mu\nu}+\frac{1}{2}\bar{S}\hg_{\mu\nu}=0 \ , \nonumber \\
& & \bar{S}=\frac{\int d^4x \sqrt{-\hg} (\frac{1}{16\pi
G}{}^{(4)}R+\mL_{matter})} {\epsilon \int d^4x\sqrt{-\hg}}
\nonumber
\\
\end{eqnarray}
that is the equation of motion found in \cite{Tseytlin:1990hn}. To
proceed to the canonical formulation we use the well know $3+1$
formalism that is the fundamental ingredient of the Hamiltonian
formalism of any theory of gravity \footnote{For recent review, see
\cite{Gourgoulhon:2007ue}.}. We consider $3+1$ dimensional manifold
$\mathcal{M}$ with the coordinates $x^\mu \ , \mu=0,\dots,3$ and
where $x^\mu=(t,\bx) \ , \bx=(x^1,x^2,x^3)$. We presume that this
space-time is endowed with the metric $\hat{g}_{\mu\nu}(x^\rho)$
with signature $(-,+,+,+)$. Suppose that $ \mathcal{M}$ can be
foliated by a family of space-like surfaces $\Sigma_t$ defined by
$t=x^0$. Let $g_{ij}, i,j=1,2,3$ denotes the metric on $\Sigma_t$
with inverse $g^{ij}$ so that $g_{ij}g^{jk}= \delta_i^k$. We further
introduce the operator $\nabla_i$ that is covariant derivative
defined with the metric $g_{ij}$.
 We also define  the lapse
function $N=1/\sqrt{-\hat{g}^{00}}$ and the shift function
$N^i=-\hat{g}^{0i}/\hat{g}^{00}$. In terms of these variables we
write the components of the metric $\hat{g}_{\mu\nu}$ as
\begin{eqnarray}
\hat{g}_{00}=-N^2+N_i g^{ij}N_j \ , \quad \hat{g}_{0i}=N_i \ , \quad
\hat{g}_{ij}=g_{ij} \ ,
\nonumber \\
\hat{g}^{00}=-\frac{1}{N^2} \ , \quad \hat{g}^{0i}=\frac{N^i}{N^2} \
, \quad \hat{g}^{ij}=g^{ij}-\frac{N^i N^j}{N^2} \ .
\nonumber \\
\end{eqnarray}
Using $3+1$
dimensional decomposition of the metric we find the primary
constraints in the form
\begin{equation}
 \pi_N\approx 0 \ , \pi_i\approx 0
\end{equation}
so that the bare Hamiltonian has the form
\begin{eqnarray}
H&=&\int d^3\bx (N(\mH_T+\epsilon B\sqrt{g})+N^i\mH_i) \ ,  \nonumber \\
\mH_T&=&\frac{16\pi G A}{\sqrt{g}}\pi^{ij}\mG_{ijkl}\pi^{kl}
-\frac{\sqrt{g}}{16\pi G A}R +\frac{A}{2\sqrt{g}}p_\Phi^2+
\frac{1}{2A}\sqrt{g}g^{ij}\partial_i\Phi\partial_j\Phi+
\frac{\sqrt{g}}{A}V(\Phi) \ ,
 \nonumber \\
\mH_i&=&-2g_{ik}\nabla_j\pi^{jk}+p_\Phi\partial_i\Phi \ ,
\nonumber \\
\end{eqnarray}
where
\begin{equation}
\mG_{ijkl}=\frac{1}{2}(g_{ik}g_{jl}+g_{il}g_{jk})-
\frac{1}{2}g_{ij}g_{kl} \ ,
\end{equation}
and where $R$ is three dimensional curvature. Further, $\pi^{ij}$
are momenta conjugate to $g_{ij}$ with non-zero Poisson bracket
\begin{equation}
\pb{g_{ij}(\bx),\pi^{kl}(\by)}=\frac{1}{2}\left(\delta_i^k\delta_j^l+
\delta_i^l\delta_j^k\right)\delta(\bx-\by) \ .
\end{equation}
Finally note that
 for simplicity we considered  the  matter Lagrangian density in the form
\begin{equation}\label{defLmatter}
\mL_{matter}=-\frac{1}{2}\hg^{\mu\nu}\partial_\mu\Phi\partial_\nu\Phi-V(\Phi)
\ .
\end{equation}
 Now the preservation of the
primary constraints imply an existence of the secondary constraints
\begin{eqnarray}
\mH'_T\equiv \mH_T+\epsilon B\sqrt{g}\approx 0 \ , \quad  \mH_i\approx 0 \  .  \nonumber \\
\end{eqnarray}
Following  the general discussion presented in previous section  we should
analyze the time evolution of all constraints. As usual it  is
 useful to introduce the smeared form of the constraints
$\mH_T,\mH_i$
\begin{equation}
\bT_T(X)=\int d^3\bx X\mH'_T \ , \quad \bT_S(X^i)=\int d^3\bx
X^i\mH_i \ ,
\end{equation}
where $X,X^i$ are smooth functions on $\Sigma$ and where these
constraints obey following Poisson bracket algebra
\begin{eqnarray}\label{pbbTS}
\pb{\bT_T(X),\bT_T(Y)}&=&\bT_S((X\partial_i Y-Y\partial_i X)g^{ij})
\
, \nonumber \\
\pb{\bT_S(X),\bT_T(Y)}&=&\bT_T(X^i\partial_iY) \ , \nonumber \\
\pb{\bT_S(X^i),\bT_S(Y^j)}&=&\bT_S(X^j\partial_j Y^i- Y^j\partial_j
X^i) \ . \nonumber \\
\end{eqnarray}
From this algebra we  see that these constraints are preserved under
the time evolution of the system. Let us then consider the action
\begin{equation}
S=\int dt d^3\bx (\pi^{ij}\partial_t g_{ij}-N(\mH_T+\epsilon
B\sqrt{g})- N^i\mH_i)+BA
\end{equation}
so that the equation of motion with respect to $A$ and $B$ have the
form
\begin{eqnarray}
A&=& \epsilon \int dt d^3\bx N\sqrt{g} \ , \quad B=\int dt d^3\bx N\frac{\delta \mH_T}{\delta A}= \nonumber \\
&=&\int dt d^3\bx N\left(\frac{16\pi
G}{\sqrt{g}}\pi^{ij}\mG_{ijkl}\pi^{kl}+ \frac{\sqrt{g}}{16\pi G
A^2}R+\frac{1}{2\sqrt{g}}p_\Phi^2-\frac{1}{2A^2} \sqrt{g}g^{ij}
\partial_i\Phi \partial_j\Phi-\frac{\sqrt{g}}{A^2}V(\Phi)\right) \ .
\nonumber \\
\end{eqnarray}
Finally inserting these values to the Hamiltonian constraint $
\mH'_T$ we  obtain the Hamiltonian form of the normalized
general relativity where the Hamiltonian constraint is manifestly
non-local but which, according to the discussion given above, has
the same constraint structure as general relativity with the
collection of the first class constraints $\pi_N\approx 0,
\pi_i\approx 0 , \mH'_T\approx 0,\mH_i\approx 0$. These constraints
could be eventually fixed by some gauge fixing procedure with
corresponding consequence for the values of the global variables $A$
and $B$.
\section{Hamiltonian Analysis of
Vacuum Energy Sequestering Model}\label{fourth}
In this section we would like to perform Hamiltonian analysis of the
model proposed in \cite{Kaloper:2013zca}. The action has the form
\begin{equation}\label{STMDVaction}
S=\int d^4x\sqrt{-\hg}\left[\frac{M_{Pl}^2}{2}{}^{(4)}R-
\Lambda-\lambda^4
\mL_{matter}(\lambda^{-2}\hg^{\mu\nu},\Phi)\right]+\sigma
\left(\frac{\Lambda}{\lambda^4\mu^4}\right) \ ,
\end{equation}
where matter couples minimally to the rescaled metric
$\tilde{g}_{\mu\nu}=\lambda^2 \hg_{\mu\nu}$. For simplicity we again
consider the matter action given in (\ref{defLmatter}). Finally
 $\Lambda$ and $\lambda$ that appear in
(\ref{STMDVaction})  are global  variables.

In the similar way as in previous section
 we find the Hamiltonian in the form
\begin{equation}
H=\int d^3\bx (N(\mH_T+\sqrt{g}\Lambda)+N^i\mH_i) \
\end{equation}
with the primary constraints of the theory
\begin{equation}
\pi_N\approx 0 \ , \pi_i\approx 0 \  ,
\end{equation}
and where
\begin{eqnarray}
\mH_T&=&\frac{2}{M_{pl}^2\sqrt{g}}
\pi^{ij}\mG_{ijkl}\pi^{kl}-\frac{M_{pl}^2}{2}\sqrt{g}R+
\frac{1}{2\lambda^2\sqrt{g}}p_\Phi^2+\frac{\lambda^2}{2}
\sqrt{g}g^{ij}\partial_i\Phi\partial_j\Phi+\lambda^4V(\Phi) \ ,
\nonumber \\
 \mH_i&=&-2g_{ik}\nabla_j\pi^{jk}+p_\Phi\partial_i\Phi \ . \nonumber \\
\end{eqnarray}
Now preservation of the primary constraints imply the secondary
constraints
\begin{eqnarray}
\partial_t\pi_N&=&\pb{\pi_N,H}=-\mH_T-\sqrt{g}\Lambda-
\equiv  -\mH'_T
\approx 0 \ , \nonumber \\
\partial_t\pi_i&=&\pb{\pi_i,H}=-\mH_i \approx 0  \ . \nonumber \\
\nonumber \\
\end{eqnarray}
Note that the smeared form of these constraints have the same
Poisson brackets as in (\ref{pbbTS}) so that given theory has the
same constraint structure.
Now the corresponding  action  has the form
\begin{equation}
S=\int dt d^3\bx (\pi^{ij}\dot{g}_{ij}-N(\mH_T+\sqrt{g}\Lambda)
-N^i\mH_i)+\sigma\left(\frac{\Lambda}{\lambda^4 \mu^4}\right)
 \ ,
 \end{equation}
 where $\Lambda,\lambda$ are determined from the equations
 \begin{eqnarray}
 \frac{1}{\mu^4\lambda^4}\sigma'&=&\int dt d^3\bx N\sqrt{g} \ ,
 \nonumber \\
4\frac{\Lambda}{\lambda^5\mu^4}\sigma'&=&-\int dt
d^3\bx(-\frac{1}{\lambda^3\sqrt{g}}p_\Phi^2+\lambda
\sqrt{g}g^{ij}\partial_i\Phi\partial_j\Phi+4\lambda^3 V(\Phi)) \ ,
\nonumber \\
 \end{eqnarray}
 where $\sigma'(x)\equiv\frac{d\sigma}{dx}$.
 As was argued in \cite{Kaloper:2013zca,Kaloper:2014dqa} the form of the function $\sigma$ should be
 determined on phenomenological grounds but in principle we can
 presume that equations given above can be solved for
 $\lambda,\Lambda$. For simplicity let us  consider the simplest
 case $\sigma(x)=x$. Then from the first equation we obtain
 \begin{equation}
 \lambda^4=\frac{1}{\mu^4\int dt d^3\bx N\sqrt{g}} \
 \end{equation}
 and from the second one we obtain
 \begin{equation}
 \Lambda=\frac{1}{4}\int dt d^3\bx
 \left(\frac{\lambda^2}{\sqrt{g}}p_\Phi^2-\lambda^6\sqrt{g}g^{ij}\partial_i
 \Phi\partial_j\Phi-4\lambda^4 V(\Phi)\right)
  \ .
  \end{equation}
  Inserting these values to the Hamiltonian we find the
  final form of the Hamiltonian formulation of this
  model that has similar canonical structure as
  the normalized general relativity.

\subsection*{Acknowledgement}
 This work was supported by the Grant
Agency of the Czech Republic under the grant P201/12/G028.

\end{document}